\documentclass[utf8]{frontiersinFPHY_FAMS} 

\usepackage{url,hyperref,lineno,microtype,subcaption}
\usepackage[onehalfspacing]{setspace}
 \usepackage[T1]{fontenc}

\def\keyFont{\fontsize{8}{11}\helveticabold }
\def\firstAuthorLast{von Neumann-Cosel {et~al.}} 
\def\Authors{Peter von Neumann-Cosel\,$^{1,2*}$ and Atsushi Tamii\,$^{3}$}
\def\Address{$^{1}$Institut f\"{u}r Kernphysik, Technische Universit\"{a}t Darmstadt, 64289 Darmstadt, Germany\\
$^{2}$Norwegian Nuclear Research Center and Department of Physics, University of Oslo, N-0316 Oslo, Norway\\
$^{3}$Research Center for Nuclear Physics, University of Osaka, Ibaraki 567-0047, Japan}
\def\corrAuthor{Peter von Neumann-Cosel}
\def\corrEmail{vnc@ikp.tu-darmstadt.de}

\begin{document}
\onecolumn
\firstpage{1}

\title[Electric dipole polarizability constraints ...]{Electric dipole polarizability constraints on  neutron skin and symmetry energy } 

\author[\firstAuthorLast ]{\Authors} 
\address{} 
\correspondance{} 

\extraAuth{}

\maketitle

\begin{abstract}

We review the experimental knowledge on the dipole polarizability (DP) of nuclei and its relation to the neutron skin thickness and properties of the neutron-rich matter equation of state (EOS). 
The discussion focuses on recent experiments using relativistic Coulomb excitation in inelastic proton scattering at extreme forward angles covering a mass range from $^{40}$Ca to $^{208}$Pb. 
Constraints on the neutron skins and the density dependence of the symmetry energy are derived from systematic comparison to calculations based on density functional theory (DFT) and ab initio methods utilizing interactions derived from chiral effective field theory ($\chi$EFT).
The results consistently favor a soft EOS around or slightly below the saturation point.
An outlook is given on possible improvements of the precision achievable in stable nuclei and studies of exotic neutron-rich unstable nuclei with upcoming experimental facilities. 

\section{}


\tiny
 \keyFont{ \section{Keywords:} dipole polarizability, neutron skin thickness, symmetry energy, density functional theory, ab initio calculations}
\end{abstract}

\section{Introduction}
\label{sec:1}

\begin{figure}
\begin{center}
\includegraphics[width=\textwidth]{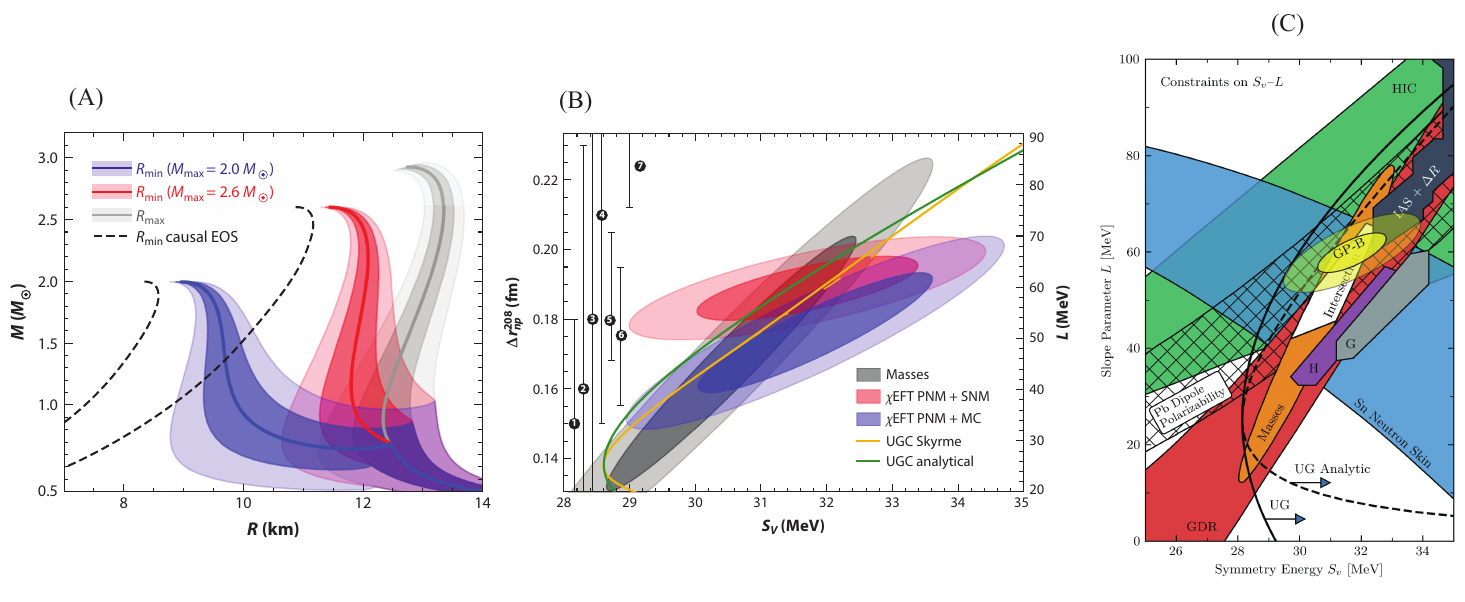}
\end{center}
\caption{ 
(A) Predictions of the mass-radius relation of neutron stars from different EOS. 
Figure taken from Ref.~\cite{lattimer21}.
(B) Theoretical constraints on the relation of $J$ (or $S_V$) and $L$. 
The points with vertical error bars on the left side represent measurements of the neutron skin thickness in $^{208}$Pb.
Figure taken from Ref.~\cite{lattimer21}, where the original references can be found.
(C) Experimental and theoretical constraints on the relation of $J$ (or $S_V$) and $L$. 
Figure taken from Ref.~\cite{drischler20}, where the original references can be found.
}
\label{fig:1}
\end{figure}

The nuclear equation of state (EOS) describes the energy per nucleon of nuclear matter as a function of proton ($\rho_p$) and neutron ($\rho_n$) densities \cite{rocamaza18}.
It governs the properties of nuclei and neutron stars \cite{lattimer21,huth22} as well as the dynamics of core-collapse supernovae \cite{yasin20} and neutron star mergers~\cite{raaijmakers21}. 
As an example, Fig.~\ref{fig:1}(A) illustrates the bounds of the mass-radius dependence of neutron stars predicted by different EOS models.   
A systematic description of the EOS from nuclear densities to those in neutron stars is a central goal of current physics \cite{koehn25}. 
Despite a wealth of new data at high densities from observations on the properties of neutron stars and neutron star mergers \cite{lattimer23} and information on the intermediate density regime from central heavy ion collisions \cite{tsang12,estee21},  experimental constraints on the EOS around the saturation density of nuclear matter $n_0 \sim 0.16$~fm$^{-1}$ are still insufficient.

The nuclear matter EOS can be approximately written as a sum of the energy per nucleon of symmetric matter and an asymmetry term
\begin{equation}
\label{eq:eos}
E(\rho,\delta) = E(\rho,\delta=0)+S(\rho)\delta^2+O(\delta^4)\,,
\end{equation}
where the nucleon density ($\rho$) and the asymmetry parameter ($\delta$) are defined by the neutron ($\rho_n$) and proton ($\rho_p$) density as
\begin{equation}
\label{eq:delta}
\rho  \equiv  \rho_n+\rho_p \, , \hspace{1cm}
\delta  \equiv  \frac{\rho_n-\rho_p}{\rho_n+\rho_p}\, .
\end{equation}
The symmetry energy factor $S(\rho)$ in Eq.~(\ref{eq:eos}) can be expanded around the saturation density $\rho_0\sim0.16$~fm$^{-3}$ as
\begin{equation}
\label{exp-sd}
S(\rho) = J+\frac{L}{3\rho_0}(\rho-\rho_0) 
+\frac{K_{\rm sym}}{18\rho_0^2}(\rho-\rho_0)^2
+\cdots\,.
\end{equation}
Here, $L$ is the slope parameter at density $\rho_0$.

The first term in Eq.~(\ref{eq:eos}) representing symmetric nuclear matter is fairly well constrained by the compressibility derived from systematic measurements of the isocalar gaint monople resonance (ISGMR) in nuclei \cite{rocamaza18}. 
Figures~\ref{fig:1}(B) \cite{lattimer21}  and (C) \cite{drischler20} illustrate the variety of experimental and theoretical constraints on $J$ (also called $S_V$ in the literature) and $L$.
While these confine possible values of $J$ to a range of about 30 to 35 MeV, the uncertainties of $L$ are much larger.

\begin{figure}
\begin{center}
\includegraphics[width=10cm]{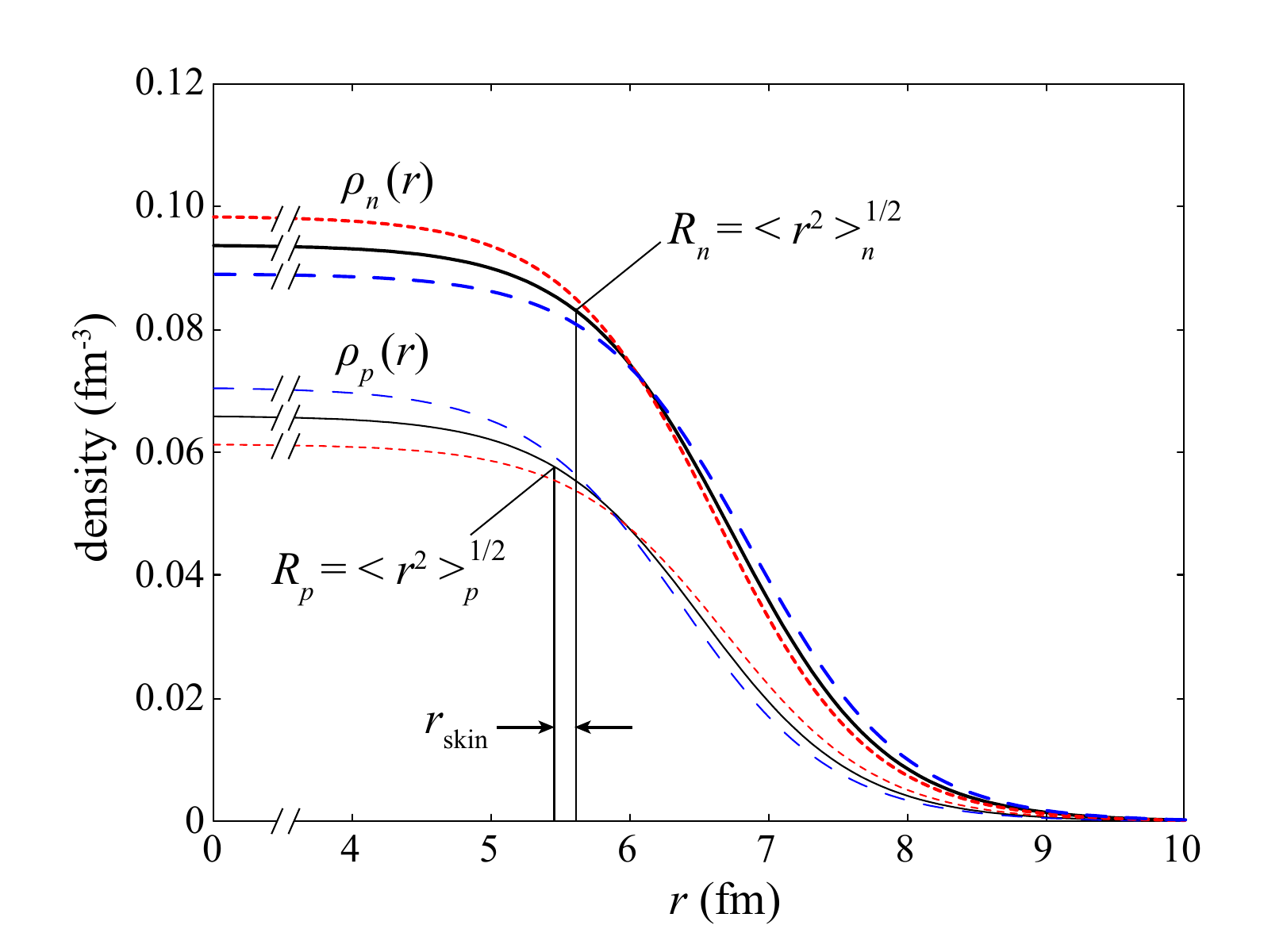}
\end{center}
\caption{ Neutron and proton density distributions are schematically shown by the thick and thin solid lines, respectively. For a larger (smaller) $L$ value, the inner density difference between neutrons and protons becomes smaller (larger), as illustrated by the blue dashed (red dotted) lines with a larger (smaller) difference at the surface, resulting in a larger (smaller) neutron skin thickness.
}
\label{fig:2}
\end{figure}

As detailed below, all relevant theoretical models predict a strong correlation of $L$ with two experimentally accessible quantities, viz.\ the neutron skin thickness and the dipole polarizability.
The connection is illustrated in Fig.~\ref{fig:2}.
The density distributions of neutrons $\rho_n(r)$ and protons $\rho_p(r)$ in the ground state of can be determined from the condition of minimum energy.
They approximately have the shape of Fermi distributions as illustrated in Fig.~\ref{fig:2} for a nucleus with $N \gg Z$.
The mean square radius of neutrons, $R_n=\left<r^2\right>^{1/2}_n$ is slightly larger than that of protons $R_p=\left<r^2\right>^{1/2}_p$.
The difference between the two, 
$r_{skin}=R_n-R_p$ 
is defined as the neutron skin thickness.

The neutron skin thickness is sensitive to the $L$ value due to the following reason.
As discussed above, the symmetry energy of nuclear matter at a given nucleon density depends on the square of the asymmetry parameter $\delta$, defined in Eq.~(\ref{eq:delta}).
The first order density dependence of the symmetry energy is represented by the slope parameter $L$.
Suppose that the density distributions in Fig.~\ref{fig:2} were determined for an $L$ value to have the minimum energy.
There are density differences between neutron and protons in the inner part (higher nucleon density) and at the surface part (lower nucleon density).
For a larger $L$ value, the density distributions change to have less density difference in the inner part thereby reducing the symmetry energy in the higher density part. 
Consequently, the neutron skin thickness and the symmetry energy at the surface becomes larger for a conserved number of neutrons and protons.

The neutron skin thickness of medium-mass and heavy nuclei has been extracted from experiments studying elastic proton scattering \cite{zenihiro10}, coherent $\pi^0$ production \cite{tarbert14}, antiprotonic atoms \cite{trzcinska01,klos07} and the isocvector (IV) spin-dipole resonance \cite{krasznahorkay99}.
Of particular importance are experiments using parity-violating polarized elastic electron scattering \cite{mammei24}.
The parity-violating part of the reaction is mediated by the weak interaction and, because of the dominance of the neutron form factor,  allows to extract the neutron density distribution in an almost model-independent way \cite{horowitz01a}. 
Such experiments have been performed for $^{208}$Pb \cite{adhikari21} and $^{48}$Ca \cite{adhikari22}. Neutron skins were determined by comparison with the well-known charge radii.  

The dipole polarizability (DP) of nuclei can be obtained from measurements of the photoabsorption cross sections.
A connection between DP, neutron skin thickness and parameters of the symmetry energy can only be made through models. 
Such calculations are presently based either on density functional theory (DFT) \cite{bender03} or ab initio coupled cluster calculations \cite{hagen14} using interactions derived from chiral effective field theory ($\chi$EFT) \cite{epelbaum09}.
Both types of models predict a strong correlation between the magnitudes of 
dipole polarizability,
$r_{\rm skin}$, and $L$.
The considerable experimental challenges of direct measurements of the neutron skin and the model dependencies of methods extracting the neutron skin from the difference of mass and charge radius \cite{thiel19} call for an alternative experimental observable.
Since properties of the symmetry energy cannot be extracted directly from experiments but require theory input, measurement of the dipole polarizability provides independent constraints.

While several experimental techniques to measure the DP are discussed, the present review mainly focuses on recent progress using relativistic Coulomb excitation in forward-angle proton scattering at energies of several hundred MeV \cite{vonneumanncosel19}.
One advantage of this method is consistent results across the neutron separation energy, while many of the other experimental techniques are limited to either the energy region below or above.
Even more important, measurements of the $E1$ strength with relativistic Coulomb excitation can be extended to exotic nuclei at rare isotope beam facilities like RIKEN, FRIB and GSI/FAIR.
Such experiments are performed in inverse kinematics, where the virtual photon flux can be boosted by using a high-$Z$ target and efficient setups with almost $4\pi$ solid angle coverage for detection of neutron emission above \cite{adrich05} and $\gamma$ emission below neutron threshold \cite{wieland09,rossi13}.
In combination with the large cross sections this will permit access to nuclei with extremely large neutron excess much closer to the properties of neutron-rich matter relevant to the physics of neutron stars.  
Besides the PUMA project \cite{aumann22} aiming at the neutron skin thickness in unstable nuclei using antiproton annihilation, dipole polarizability measurements with relativistic Coulomb exitation are probably the only experimental probe promising insight into properties of the symmetry energy over a wide range of neutron-to-proton ratios.

The paper is organized as follows.
Section \ref{sec:2} discusses how information on the neutron skin thickness and symmetry energy can be inferred from model calculations based on DFT (Sect.~\ref{sec:21}) and ab initio methods (\ref{sec:22}). 
Section \ref{sec:3} is devoted to experimental issues.
A short discussion of the available techniques in Sec.~\ref{sec:31} is followed by a description of methods to disentangle electric and magnetic contributions to the DP in Sec.~\ref{sec:32}. 
The relevance of experimental information in the energy region of the IVGDR as well as below the neutron threshold and above the IVGDR is compared in Secs.~\ref{sec:33}-\ref{sec:35}.
The comparison of experimental and theoretical results (Sec.~\ref{sec:4}) for a range of nuclei from $^{40}$Ca to $^{208}$Pb and constraints on neutron skin thickness and paramaters of the symmetry energy extracted thereof are presented in Sec.~\ref{sec:41} for DFT and Sec.~\ref{sec:42} for ab initio approaches.
Section \ref{sec:43} focuses on the difficulties to simultaneously describe results of parity-violating elastic electron scattering and DP experiments with present-day models.
Finally, Sec.~\ref{sec:44} discusses systematics of the DP and the role of volume and surface contributions to the symmetry energy.
A summary and an outlook are given in Sec.~\ref{sec:5}.

\section{Relation between dipole polarizability, neutron skin thickness and symmetry energy}
\label{sec:2}

In this section we discuss how information on the neutron skin thickness and parameters of the symmetry energy can be inferred from the comparison of the experimental dipole polarizability to theoretical predictions. 
At the moment, there are two classes of models either based on DFT or an ab initio coupled cluster approach.
Since isovector observables are not well constrained in DFT, quantitative predictions of the DP can vary considerably.
However, one can establish a robust correlation between the parameters $J$ and $L$ of the symmetry energy through $\alpha_D$.
With ab initio based models one aims at an absolute prediction of $\alpha_D$ and the underlying symmetry energy parameters of the interaction can be used to calculate the  EOS. 

The dipole polarizability $\alpha_{D}$ is related to the reduced $B(E1)$ transition strengths and the photoabsorption cross sections $\sigma_{\rm abs}$ by
\begin{equation}
\label{eq:DP}
  \alpha_D
  =
  \frac{\hbar c}{2\pi^{2}}  
  \int_0^\infty \frac{\sigma_{abs}}{E_{x}^{2}}{\rm d}E_{x} 
  = 
  \frac{8 \pi}{9} \int_0^\infty \frac{B(E1)}{E_{x}}{\rm d}E_{x}.
\end{equation}
While the integral runs to infinity in principle, because of the inverse energy weighting a measurement of the $E1$ strength up to excitation energies of about 60 MeV in light \cite{fearick23} or 30 MeV in heavy nuclei \cite{hashimoto15} is sufficient to achieve saturation. 
Thus, $\alpha_ {\rm D}$ is dominated by the isovector giant dipole resonance (IVGDR) but contributions from the energy regions below and above are non-negligible as discussed in Sec.~\ref{sec:3}.

\subsection{Connections in density functional theory}
\label{sec:21}

\begin{figure}
\begin{center}
\includegraphics[width=12cm]{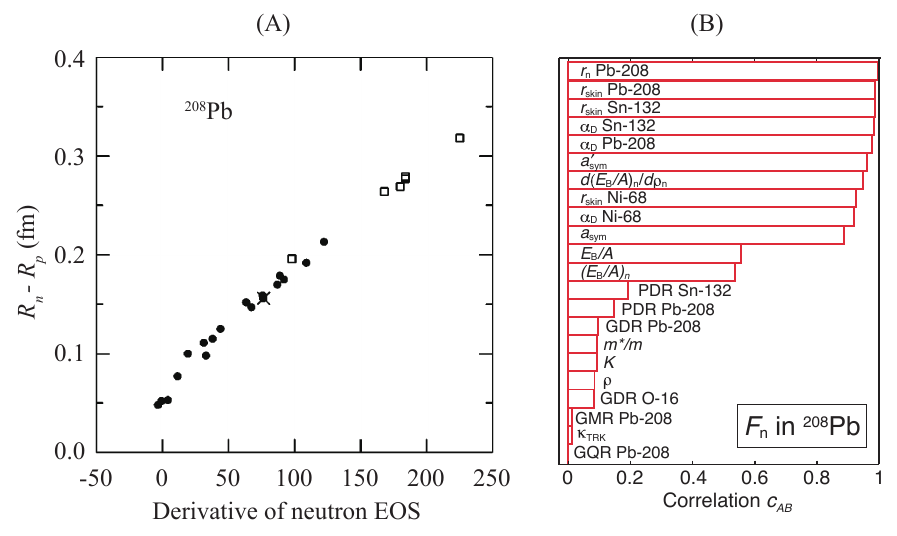}
\end{center}
\caption{ 
(A) Correlation between the neutron skin thickness in $^{208}$Pb and $L$ for a large set of DFT interactions. 
Figure taken from Ref.~\cite{typel01}. 
(B) Correlation of various observables in $^{208}$Pb with the neutron form factor at $q=0.45$~fm$^{-1}$.
Figure taken from Ref.~\cite{reinhard10}.
}
\label{fig:3}
\end{figure}
An approximately linear correlation between $r_{skin}$ and $L$ was demonstrated in Hartee-Fock calculations of $^{208}$Pb with relativistic \cite{typel01} and Skyrme \cite{brown00} density functionals as illustrated in Fig.~\ref{fig:3}(A).
A comprehensive investigation of correlations between IV experimental observables and bulk parameters of DFT models \cite{reinhard10} demonstrates that these two quantities are also correlated with $\alpha_{\rm D}$ in heavy nuclei.
 Figure~\ref{fig:3}(B) shows, as an example, the correlations with the neutron form factor of $^{208}$Pb, which can be derived from a parity-violating elastic electron scattering experiment.

\begin{figure}
\begin{center}
\includegraphics[width=14cm]{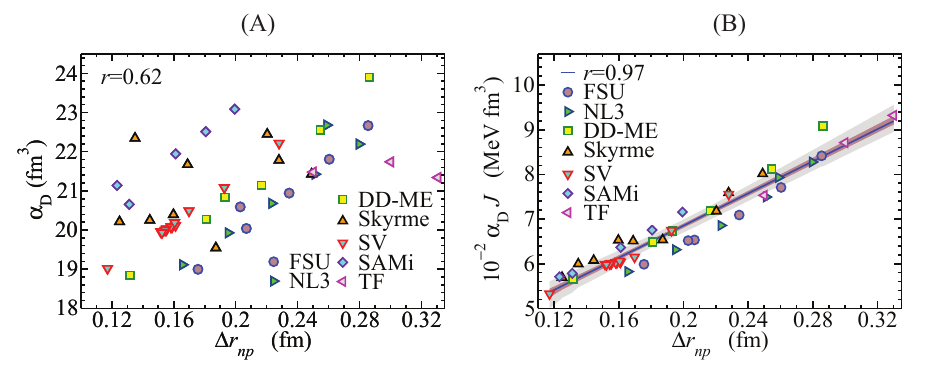}
\end{center}
\caption{
(A) Dipole polarizability against neutron skin thickness in 
$^{208}$Pb Pb predicted by modern DFT interactions.  
(B) Same for dipole polarizability times symmetry energy at saturation density.
The results are well described by a linear fit.
Figures taken from Ref.~\cite{rocamaza13}, where the original references for the various interactions can be found.
}
\label{fig:4}
\end{figure}

While this type of correlations is observed for all interactions, absolute values show large differences. 
In general, the magnitude of IV quantities like $\alpha_{\rm D}$ is not well constrained in DFT models, since the model parameters are typically fitted to binding energies and charge radii of selected nuclei, which show little sensitivity to IV parts of the nuclear interaction. 
A study of the relation between $\alpha_{\rm D}$ and the neutron skin in $^{208}$Pb for a large number of interactions illustrates the problem \cite{rocamaza13}.
Figure~\ref{fig:4}(A) shows that the predictions for the neutron skin vary from 0.12 to 0.32 fm, and for a given value of $r_{\rm skin}$ predictions for $\alpha_{\rm D}$ scatter wildly.
However, the product of $\alpha_{\rm D} \times J$ plotted versus $r_{\rm skin}$ (or $L$) shows a linear dependence with a high correlation coefficient \cite{rocamaza13}, cf.\ Fig.~\ref{fig:4}(B).
This relation can be understood within the droplet model \cite{myers74} and provides a correlated range of $J,L$ values as indicated for the case of $^{208}$Pb \cite{tamii11} in Fig.~\ref{fig:1}(C).

\subsection{Connections in ab initio models}
\label{sec:22}

\begin{figure}[b]
\begin{center}
\includegraphics[width=\textwidth]{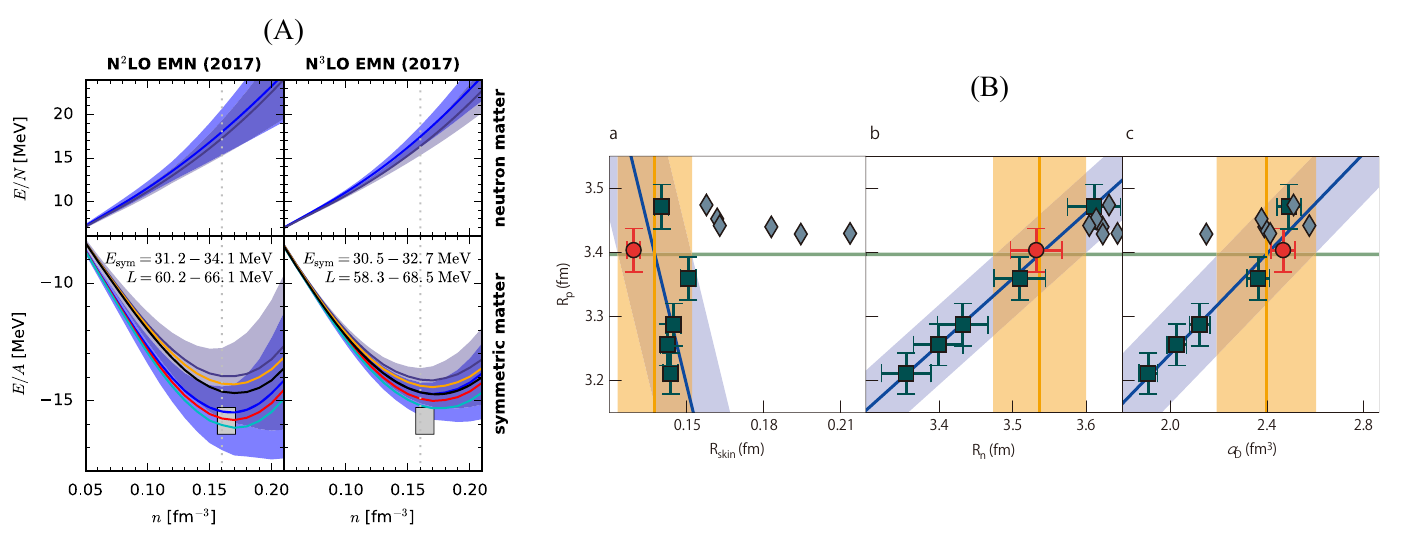}
\end{center}
\caption{
(A) Energy per particle in neutron matter (top row) and symmetric nuclear matter (bottom row) based on chiral interactions at 
N$^2$LO  (first column) and N$^3$LO (second column) fit to the empirical saturation region (grey box). 
The blue and gray bands estimate the theoretical uncertainty assuming different parameter contraints. 
Figure taken from Ref.~\cite{drischler19}.
(B) Predictions of the neutron skin (a), point-neutron radius  (b) and electric dipole polarizability (c) versus the point-proton radius
for $^{48}$Ca.
Ab initio results with the NNLO$_{\rm sat}$ interaction \cite{ekstroem15} and chiral interactions \cite{hebeler11} are shown as red circles and squares, respectively,
The blue line represents linear fits to the ab initio predictions. The horizontal green line marks the experimental value of the $^{48}$Ca charge radius. 
Its intersection with the blue lines and the blue bands yields the vertical orange lines and orange bands, respectively, giving the predicted range for the ordinates.
Figure taken from Ref.~\cite{hagen16}, where the original references of the shown DFT interactions can be found.
}
\label{fig:5}
\end{figure}

Ab initio calculations based on interactions derived from $\chi$EFT play an important role in the attempt to systematically describe the EOS of neutron-rich matter at all densities \cite{huth22}.
Figure~\ref{fig:5}(A) displays examples of next-to-next-to-next-to-leading order predictions of the density behavior in the nuclear regime \cite{drischler19}.
The upper and lower parts present the neutron and symmetric matter results, respectively, for two different families of interactions with somewhat different symmetry energy parameters shown in the left and right colums. 
The gray boxes indicate the value of the saturation density.
The colored curves correspond to different cutoff paramaters of the model space; for details see Ref.~\cite{drischler19}.

Predictions of $\alpha_{D}$ and correlations with proton and neutron radii based on $\chi$EFT interactions have been obtained from calculations based on a coupled-cluster expansion of the wave functions \cite{hagen14} combined with the Lorentz-integral-transform approach to extract the $E1$ strength \cite{bacca14}. 
An example of such calculations for $^{48}$Ca \cite{hagen16} is presented in Fig.~\ref{fig:5}(B), where the correlation of $r_{p}$ with $r_{n}$, $r_{kin}$ and $\alpha_{D}$ is displayed together with representative examples of DFT predictions.  
While the DFT results predict neutron skin values ranging from 0.16 fm to 0.22 fm, the ab initio results based on a set of interactions from Refs.~\cite{hebeler11,ekstroem15} consistently predict rather small values varying from 0.12 to 0.15 fm.

A major difference between the two theoretical approaches lies in the predicted relation between proton and neutron radius.
The DFT predictions of $r_{p}$ are approximately constant, most likely because the charge radius of $^{48}$Ca is in all cases part of the data set used to fix the model parameters. 
The ab initio calculations, on the other hand, predict a linear correlation, leading to the approximate constancy of the neutron skin.
The absolute value of $\alpha_{D}$ in the ab initio models shows a larger variation compared to the DFT calculations, but can be well described by a linear correlation similar to $r_{n}$. 
As discussed in the following, these correlations permit for an extraction of constraints on the range of symmetry energy parameters based on the successful description of experimentally measured polarizabilities and charge radii.
This type of calculation has been limited so far to closed-(sub)shell nuclei. 
For recent attempts of an extension to open-shell nuclei see Refs.~\cite{bonaiti24a,brandherm25}.

\section{Dipole polarizability from experiment} 
\label{sec:3}

In this section, we discuss the experimental methods to extract the $B(E1)$ distribution in nuclei and the DP.
It is technically difficult to directly measure the DP of nuclei as the response to a static electric field although there exist exceptional cases of very light nuclei; see e.g.\ the works studying the deviation of elastic scattering cross section from Rutherford scattering~\cite{rodning82,lynch82}.
Instead, $B(E1)$ or $\sigma_{abs}$ distributions are measured and integrated to determine the DP by Eq.~(\ref{eq:DP}).
Some of the experimental methods discussed below are restricted in the accessible excitation energy range, i.e., the techniques are applicable below or above neutron emission threshold ($S_n$) only.
Thus, the role of contributions to the DP below $S_n$, from the IVGDR and from the energy region above the IVGDR are discussed in more detail.
Both $E1$ and $M1$ transitions are excited and possible ways of their distinction are briefly presented.  

\subsection{Experimental Methods}
\label{sec:31}

\subsubsection{Photoneutron measurement}
\label{sec:311}

\begin{figure}
\begin{center}
\includegraphics[width=8cm]{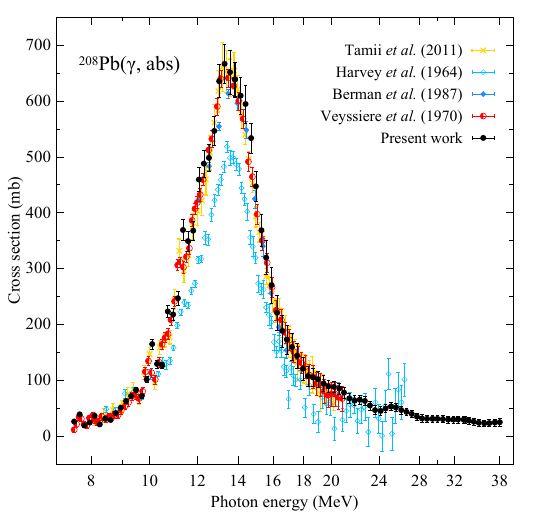}
\end{center}
\caption{
Comparison of photoabsorption cross sections of $^{208}$Pb from different experments. 
Figure taken from Ref.~\cite{gheorghe24}, where the original references can be found.
}
\label{fig:6}
\end{figure}

The photoexcitation of nuclei above the neutron separation energy was intensively studied by using photo-neutron measurement.
The photoneutron reaction is conventionally written as $(\gamma,xn)$, where $x$ stands for the number of emitted neutrons after photo-excitation.
The $(\gamma,xn)$ cross section is the sum of the $(\gamma,1n)$, $(\gamma,2n)$, ... + $(\gamma,np)$ above the respective thresholds.
From 1960's to 80's, positron annihilation in flight was used for producing a $\gamma$-ray beam at Lawrence Livermore National Laboratory (LLNL) and at Saclay.
Neutrons emitted after interaction with a target were thermalized and detected.
For details see Refs.~\cite{berman75,dietrich88}.

Neutron emission is the dominant decay process after photoexcitation for a nucleus as heavy as $^{208}{\rm Pb}$ since charged particle decays are strongly suppressed by the Coulomb barrier and the $\gamma$ decay branch is as low as 1-2\%~\cite{beene90}.
Thus, the $(\gamma,xn)$ cross sections in heavy nuclei can be compared with total photoabsorption cross sections.
The $^{208}{\rm Pb}(\gamma,xn)$ cross sections are plotted in Fig.~\ref{fig:6}.
The data were taken at LLNL (light blue open circles~\cite{harvey64} and blue solid circles~\cite{berman87}) and at Saclay (red half-filled circles~\cite{veyssiere70}).
The results from the two laboratories show clear discrepancies, which is also true for some other nuclei. 
Kawano et al.~\cite{kawano20} reported that "in general, the Saclay $(\gamma,n)$ cross sections are larger than the Livermore data, whereas the Saclay $(\gamma,2n)$  cross sections are smaller than the corresponding Livermore data".

Later, quasi-monoenergetic photon beam produced by laser Compton backscattering (LCBS) became available
at the National Institute of Advanced Industrial Science and Technology (AIST)~\cite{toyokawa09},
the High Intensity $\gamma$-ray Source (HI$\gamma$S) facility at the Triangle University National Laboratory~\cite{weller09}
and the NewSUBARU facility ~\cite{amano09,horikawa10}.
An electron beam in a storage ring is irradiated by laser photons to produce high-energy photons by head-on collisions~\cite{kawano20}. 
The scattered photons are collimated to have a narrow energy distribution. 
The photon energy is variable either by changing the electron beam energy or the laser frequency.
The $^{208}{\rm Pb}(\gamma,xn)$ cross section data measured at NewSUBARU~\cite{gheorghe24} are plotted as solid black circles in Fig.~\ref{fig:6}.


\subsubsection{Total photoabsorption}
\label{sec:312}

Total photon absorption was studied by applying transmission measurements. 
In this method, the attenuation of photons in a thick target was measured as a function of the photon energy for extraction of the photoabsorption cross sections.
At the Mainz electron accelerator, a narrow photon beam was produced by Bremsstrahlung of an electron beam.
The average photon flux was $10^9$ photons/MeV at 20 MeV. 
Two identical Compton spectrometers monitored the photon flux before and after a natural abundance target with a thickness of 40-200~cm~\cite{ahrens75}.
The dominant atomic photoabsorption cross sections needed to be subtracted.
A high-resolution transmission measurement at AIST was reported for $^{28}{\rm Si}$ using a HPGe detector~\cite{harada01}.
Recently an experimental setup  for photon transmission measurement is under operation at the photon tagger NEPTUN~\cite{savran10} at the S-DALINAC accelerator in Darmstadt.

\subsubsection{Compton scattering}
\label{sec:313}

Compton scattering from $^{208}{\rm Pb}$ was measured at Mainz using quasi-monoenergetic photons produced by positron anihilation in flight~\cite{schelhaas88} up to a photon energy of 143 MeV.
The flux of the photon beam was monitored with a Compton spectrometer. 
Elastically scattered photons were detected by large-volume  NaI scintillation counters.
The multipolarity dependent cross sections were analyzed using the angular distributions.
The imaginary part of the scattering cross sections at zero degrees is related to the total photon cross section.

\subsubsection{Bremsstrahlung excitation functions}
\label{sec:314}

Photonuclear cross sections have been extracted from the radioactive decay of residual nuclei populated in particle emission after irradiation with thick-target bremstrahlung. 
The excitation-energy dependence can be determined by variation of the bremstrahlung endpoint energy with an unfolding procedure \cite{penfold59}.
However, this requires precise knowledge of the bremstrahlung spectra experimentally not available.
While such spectra can be reliably calculated \cite{rasulova25} with present-day Monte Carlo codes such as GEANT4 \cite{agostinelli03}, older versions contained poor approximations \cite{vonneumanncosel94}.
Results deduced from phenomenological approximations or using the analytical description of thin-target bremsstrahlung have potentially very large systematic uncertainties, typically not included in the quoted errors. 

\subsubsection{Relativistic Coulomb excitation}
\label{sec:315}

Relativistic Coulomb excitation is an important experimental tool to study the electric dipole response at RIB facilities.
At beam energies of several hundred MeV/nucleon, cross sections are large and cover an excitation energy range including the IVGDR.
The small number of beam particles can be compensated for neutron-rich nuclei by placing a neutron detector under $0^\circ$, since at highly relativistic velocities a small angular opening is sufficient to cover the full 4$\pi$ solid angle range in the center-of-mass system.
The method has been applied to study e.g.\ halo nuclei \cite{nakamura06} and neutron-rich oxygen isotopes \cite{leistenschneider01}. 
DP measurements in heavier nuclei have been performed for $^{68}$Ni \cite{wieland09,rossi13} and $^{130,132}$Sn \cite{adrich05}.

The method has also been developed to study stable nuclei using inelastic proton scattering under extreme forward angles including $0^\circ$.
Such experiments require advanced methods to remove the background from beam halo and atomic small-angle scattering in the target. 
Zero-degree setups have been realized at the Research Center for Nuclear Physics, Osaka, Japan for proton energies up to 400 MeV \cite{tamii09} and at the iThemba Laboratory for Accelerator-Based Science, Faure, South Africa for 200 MeV \cite{neveling11}. 
An overview of experiments, data analysis and physics problems addressed is provided in Ref.~\cite{vonneumanncosel19}.

\subsubsection{Nuclear resonance fluorescence}
\label{sec:316}

Nuclear resonance fluorescence (NRF) or $(\gamma,\gamma^\prime)$ experiments study the $\gamma$ emission after resonant absorption of a photon.
The reaction selectively excites states with large ground-state braching ratios.
The cross section contributions due to the decay to excited states can be estimated from the population of the lowest excited states.
The experiments can be performed with Ge detectors and thus offer unique energy resolution. 
The measured quantities depend on the product of photo-absorption cross sections and ground-state branching ratio, thus the method is limited to excitation energies below the neutron threshold because of the dominance of particle decay widths in the continuum.
Experimental methods, physics and applications are discussed in a recent review \cite{zilges22}.

\subsection{Decomposition of $E1$ and $M1$ contributions}
\label{sec:32}

A general problem of all experimental methods discussed above is the removal of magnetic contributions to the photoabsorption cross sections and the derived DP.
Overall, contributions of $M1$ strength to the DP are small except for very light nuclei \cite{knuepfer81}. 
However, they become relevant in the excitation energy region of the spinflip $M1$ resonance \cite{heyde10}.
Therefore, techniques to decompose $E1$ and $M1$ contributions are important. 

No $E1/M1$ decomposition can be performed for the photo-neutron and total photo-absorption experiments.
In the excitaton energy regime relevant to determine the DP, they can be distinguished in Compton scattering by combining measurements at forward and backward angles.
The multipolarity can also be determined in NRF experiments using transversely polarized photons \cite{pietralla24}.
Measurements of the response relative to the polarization plane permit an unique assignment of the electric or magnetic character of the emitted radiation.
Polarized beam can be extracted from off-axis bremsstrahlung or LCBS. 
The latter method is particularly efficient because the polarization of the laser light is fully transferred to the photon beam \cite{zilges22}.

\begin{figure}
\begin{center}
\includegraphics[width=15cm]{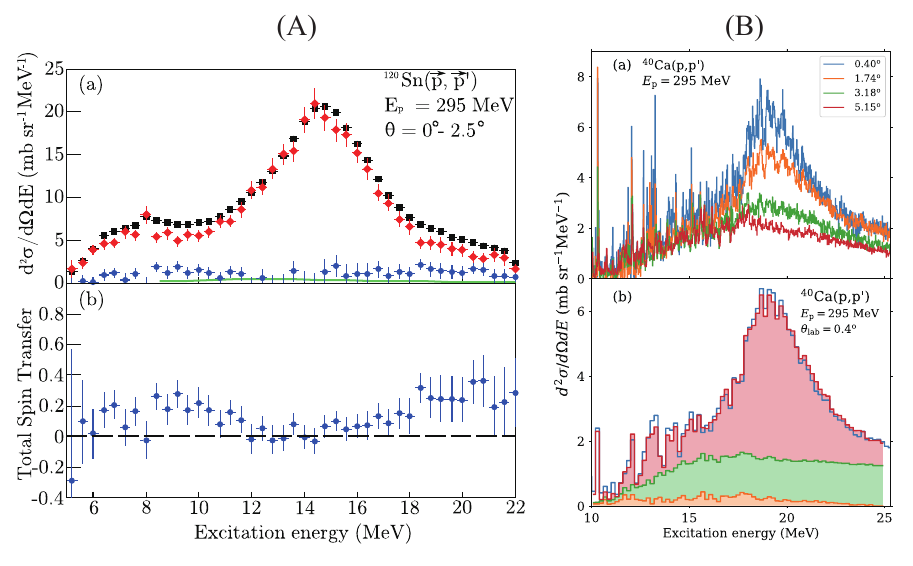}
\end{center}
\caption{
(A) Top: Double differential cross sections of the $^{120}$Sn($\vec{p}$,$\vec{p}^\prime$) reaction (black squares) and decomposition into non-spinflip (red diamonds) and spinflip (blue circles) parts. 
The green solid line shows the cross sections due to excitation of the ISGQR. 
Bottom: Total spin transfer as defined in Ref.~\cite{suzuki00}.
Figure taken from Ref.~\cite{hashimoto15}.
(B) Top: Spectra of the $^{40}$Ca$(p,p^\prime)$ reaction at 
$E_{p} = 295$ MeV and different scattering angles.
Bottom: Example of the decomposition of the spectrum at $\Theta_{lab} = 0.4^\circ$ into contributions of $L > 1$ multipoles (orange), continuum background (green) and $E1$ (red).
Figure taken from Ref.~\cite{fearick23}.
}
\label{fig:7}
\end{figure}

In relativistic Coulomb excitation, the virtual photon spectrum in forward direction is dominated by $E1$.
However, in the proton scattering experiments close to $0^\circ$ one has to consider contributions to the cross sections due to nuclear excitaton of the spinflip $M1$ strength.
Two independent methods have been applied to separate $E1$ and $M1$ cross section parts based either on the total spin transfer derived from the combined information of polarization transfer observables or from a multipole decomposition analysis (MDA) of the cross section angular distributions \cite{vonneumanncosel19}.
Figure~\ref{fig:7} presents some illustrative examples.
The lower panel of Fig.~\ref{fig:7}(A) displays the total spin transfer at $0^\circ$ for the nucleus $^{120}$Sn  \cite{hashimoto15} derived from measurements of the polarization-transfer observables $D_{LL}$ and $D_{SS}$ \cite{suzuki00}.
The upper panel presents the differential cross sections at $0^\circ$
and their decomposition in non-spinflip ($E1$ from Coulomb  excitation) and spinflip ($M1$ from nuclear excitation) parts. 

An example of the MDA analysis is presented in Fig.~\ref{fig:7}(B) for $^{40}$Ca \cite{fearick23}.
Spectra at different scattering angles are displayed in the upper panel demonstrating strongly forward-peaked cross sections in the energy region of the IVGDR expected for Coulomb excitation.
The lower panel shows the partial contributions to the cross sectios at the most forward angle measured resulting from the MDA: $E1$ (red), multipoles $L > 1$ (orange) and nuclear background from quasifree scattering (green).
Note that $M1$ strength was neglected for $^{40}$Ca because it is concentrated in a single state \cite{gross79}.
A comparison of the two independent methods was made in studies of $^{96}$Mo \cite{martin17}, $^{120}$Sn \cite{hashimoto15} and $^{208}$Pb \cite{tamii11} and good correspondence of the resulting $E1$ and $M1$ cross sections was found. 
The results shown in Fig.~\ref{fig:7}(A) have also been confirmed in a MDA analysis \cite{bassauer20a}.
Since the polarization transfer measurements require secondary scattering, statistics are limited.
Thus, in most $(p,p^\prime)$ experiments the $E1/M1$ decomposition was restricted to MDA.

\subsection{Contributions from the IVGDR}
\label{sec:33}

\begin{figure}
\begin{center}
\includegraphics[width=\textwidth]{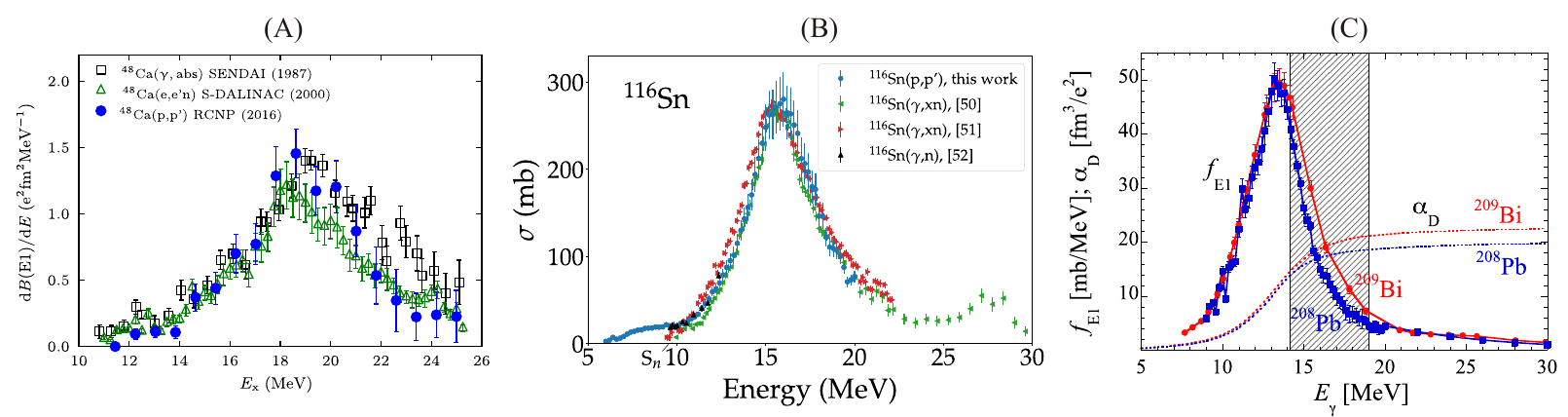}
\end{center}
\caption{Comparison of photoabsorption cross sections from different experiments.
(A) $B(E1)$ strength distributions in $^{48}$Ca.
(B) Photoabsorption cross sections in $^{116}$Sn.
(C) Photon strength functions (solid lines) of $^{208}$Pb (blue squares) and $^{209}$Bi (red circles) and the corresponding estimate of the contribution to the DP (dashed lines).
Figures taken from (A) Ref.~\cite{birkhan17}, (B) Ref.~\cite{bassauer20} and (C) Ref.~\cite{goriely20}, where the original references can be found.
}
\label{fig:8}
\end{figure}

The largest contribution to the DP stems from the IVGDR, whose energy centroids lie well above $S_n$.
It is experimentally accessible with different techniques, and the comparison of results for the same nucleus provides an estimate of the typical accuracy of the DP.
One can also average over results obtained with independent methods, thereby reducing the error bars. 
Some illustrative examples are presented in Fig.~\ref{fig:8}(A) and (B) for $^{48}$Ca and $^{116}$Sn, respectively.

The $B(E1)$ strength distributions in $^{48}$Ca obtained from $(p,p^\prime)$  \cite{birkhan17} (blue circles) and $(e,e^\prime n)$ \cite{strauch00} (green triangles) agree well.
Results derived from the bremsstrahlung-induced activity of $^{47}$Ca \cite{okeefe87} agree on the low-energy wing of the resonance but are significantly larger than the other data on the high-energy side.
This can probably be traced back to the problems discussed in Sec.~\ref{sec:314}.
The second example (B) compares photoabsorption cross sections for $^{116}$Sn from relativistic Coulomb excitation \cite{bassauer20} (blue squares) with $(\gamma,xn)$ data \cite{fultz69,lepretre74} (green left and red right arrows).
Reasonable agreement is observed in the maximum region of the IVGDR, but one ﬁnds signiﬁcant differences on the low-energy flank.
Such deviations are systematically observed in the stable Sn isotope chain, and for some isotopes also at the high-energy flank \cite{bassauer20a}.

In general, studies of the $(\gamma,n)$ and $(\gamma,xn)$ reactions with LCB beams at NewSUBARU ageee well the $(p,p^\prime)$ results from RCNP, see e.g\ Ref.~\cite{utsunomiya09} for a study of Sn isotopes (black upward arrows in Fig.~\ref{fig:8}(B)) or for $^{208}$Pb \cite{gheorghe24} discussed in Sec.~\ref{sec:311}.
However, a puzzling result has been reported for $^{209}$Bi shown in Fig.~\ref{fig:8}(C).
Although it differs from $^{208}$Pb by one extra neutron only, additional strength is seen on the high-energy side of the IVGDR leading to a difference in $\alpha_D$ not predicted by any model.
This particular case certainly needs further investigation.

\subsection{Contributions from the PDR}
\label{sec:34}

All particle-emission coincidence experiments accessing the $E1$ strength are limited to the excitation region above the lowest particle separation threshold.
Experimental evidence has accumulated that in nuclei with neutron excess significant $E1$ strength -- often concentrated in a resonance-like structure commonly termed PDR -- can be found below \cite{savran13,bracco19}.
Low-energy $E1$ strength is also found in lighter nuclei with $N \approx Z$.
Its contribution to the DP can be significant because of the inverse energy weighting, cf.\ Eq.~(\ref{eq:DP}).
As examples, they amount to about 10\% in $^{58}$Ni \cite{brandherm25} and 8-13 \% in the stable Sn isotopes \cite{bassauer20}. 

\begin{figure}
\begin{center}
\includegraphics[width=8cm]{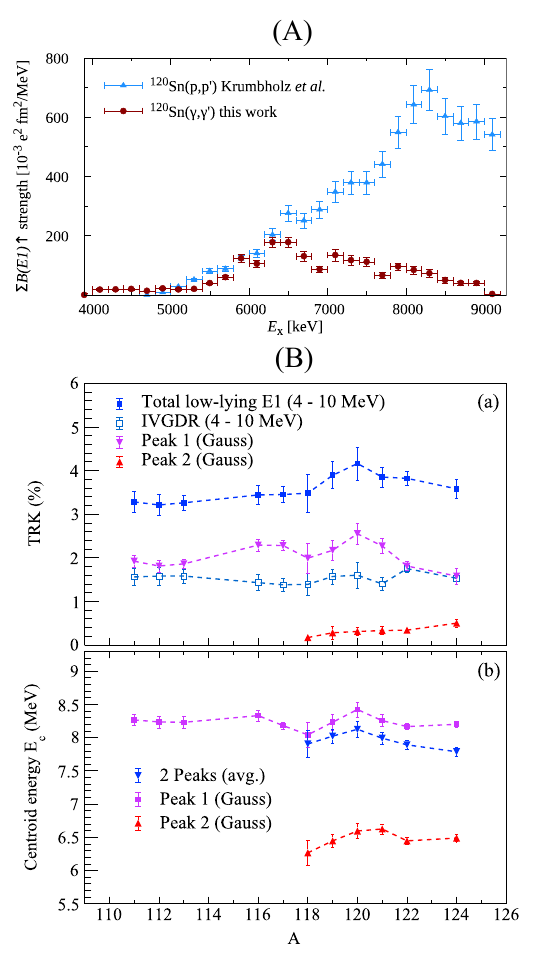}
\end{center}
\caption{
(A) Comparison of $B(E1)$ strength distributions in $^{120}$Sn from resolved states in a NRF experiment (red circles) \cite{muescher20} from the (p,p$^\prime)$ reaction (blue triangles) \cite{krumbholz15}.
Figure taken from Ref.~\cite{muescher20}.
(B) Systematics of the total electric dipole strength in $^{111-124}$Sn integrated over the energy region $4-10$ MeV and its decomposition into contributions from the tail of the IVGDR and one or two (for masses $\geq 118$) resonances. 
Top: Strengths in \% of the TRK sum rule. Bottom: Centroid energies.
Figure taken from Ref.~\cite{markova25}. 
}
\label{fig:9}
\end{figure}

The majority of data on low-energy $E1$ strength stems from $(\gamma,\gamma^\prime)$ experiments \cite{zilges22}.
They suffer from the problem that branching ratios to excited states are typically unknown and the extracted strength based on the g.s.\ transitions represents a lower limit only.
Taking $^{120}$Sn as an example, the resulting $B(E1)$ strength distribution \cite{muescher20} reasonably agrees with a ($p,p^\prime$) experiment \cite{krumbholz15} measuring the total excitation strength up to about 6.5 MeV but totally underestimates the strength at higher excitation energies, cf.\ Fig.~\ref{fig:9}(A).
Attempts have been made to model the inelastic contributions assuming statistical decay (see e.g.\ \cite{rusev09}) but tend to overestimate contributions at very low excitation energies.
However, progress has been made recently by analyzing the cumulative population of the first $2^+$ state in even-even nuclei \cite{zilges22}.
For the quoted example $^{120}$Sn, good agreement between the two experimental methods is achieved \cite{muescher20}.

The origin of the low-energy $E1$ strength in nuclei with neutron excess is a topic of current debate.
It has been suggested to arise from an oscillation of the excess neutrons forming a skin against the (approximately) isospin-saturated core \cite{paar07,lanza23}.
If true, its strength should be related to the neutron skin thickness and in turn to the parameters of the symmetry energy \cite{klimkiewicz07,carbone10,bertulani19}.
However, a recent study of the Sn isotope chain for mass numbers 111-124 casts doubts on such a picture.
The correlation between neutron excess and neutron skin thickness in Sn isotopes has been experimentally demonstrated with different methods \cite{krasznahorkay04}, but based on combined data from Oslo \cite{markova21,markova22,markova23} and ($p,p^\prime$) experiments \cite{bassauer20a} only a minor fraction of the photoabsorption cross section (expressed as fraction of the TRK sum rule) can be related to the PDR \cite{markova24}.
A decomposition into the tail of the IVGDR and two resonance-like structures is shown in Fig.~\ref{fig:9}(B) \cite{markova25}. 
The contribution interpreted as PDR is much smaller than those of the IVGDR and the prominent structure around 8 MeV.
These findings rather point to an interpretation of the PDR as low-energy part of a toroidal $E1$ mode \cite{repko19,vonneumanncosel24}.
At present, understanding the nature of the PDR remains an open problem.
It is clear, however, that DFT predictions restricted to 1p-1h excitations cannot reliably estimate the low-energy $E1$ strength distribution for cases where data are unavailable \cite{markova25}.

\subsection{Contributions from high excitation energies}
\label{sec:35}

At excitation energies beyond the giant resonance region photonuclear cross sections typically contribute a few percent only to the DP. 
However, for precision results they need to be considered. 
Data up the pion threshold have been measured for a few cases, viz.\ $^{\rm nat}$Ca \cite{ahrens75}, $^{\rm nat}$Sn \cite{lepretre81} and $^{208}$Pb \cite{veyssiere70,schelhaas88}.
They show approximately constant cross sections as function of excitation energy and were considered for the extraction of the DP from $(p,p^\prime)$ experiments \cite{tamii11,hashimoto15} neglecting an isotopic dependence.
The dominant excitation mechanism in this energy regime is the quasideuteron effect \cite{levinger51}.
It has been pointed out by Roca-Maza et al.~\cite{rocamaza15} that these contributions are not included in model calculations based on DFT and should thus be removed in comparison to theoretical predictions. 
For heavy nuclei they can be estimated using Ref.~\cite{chadwick91}, while in light nuclei they are negligible in the energy range covered by the models \cite{birkhan17,fearick23,brandherm25}.

The ratio of Coulomb excitation to quasifreee cross sections in the ($p,p^\prime$) experiments \cite{vonneumanncosel19} drops with decreasing mass number limiting in some cases the excitation energy range accessible with a MDA for the extraction of $E1$ cross sections.
In such cases model-dependent corrections have to be applied.
In the study of the Sn isotopic chain \cite{bassauer20}, these were based on QRPA calculations folded with a Lorentzian to reproduce the experimentally measured width of the IVGDR.
A particularly promising approach is discussed in Ref.~\cite{brandherm25} for the example of $^{58}$Ni.
An extension of the QRPA calculations to include quasiparticle vibration coupling has been successful in describing the width of the ISGMR and curing a longstanding discrepancy between the compressibility values extracted from $^{208}$Pb and lighter nuclei \cite{li23,litvinova23}. 
The application to $^{58}$Ni demonstrates that the predicted high-energy tail of the IVGDR is largely independent of the chosen interaction \cite{brandherm25}.
This can be understood to result from the dominance of stochastic coupling \cite{vonneumanncosel19a}, i.e., the strength distribution is mainly determined by the density of states and an average coupling matrix element between the 1p-1h and more complex states.


\section{Extracting neutron skin thickness and symmetry energy properties from dipole polarizability data}
\label{sec:4}

In this section we discuss constraints on the neutron skin thickness and symmetry energy properties derived from the comparison between model predictions and experimental studies of the DP.
These refer to specific nuclei like $^{40}$Ca, $^{48}$Ca and $^{208}$Pb but also systematic isotopic trends or a global mass dependence.
The difficulties of presently available models to simultaneously account for measured polarizabilities and asymmetries in parity-violating elastic electron scattering are illuminated. 

\subsection{Constraints based on density functional theory}
\label{sec:41}

\begin{figure}[h!]
\begin{center}
\includegraphics[width=18cm]{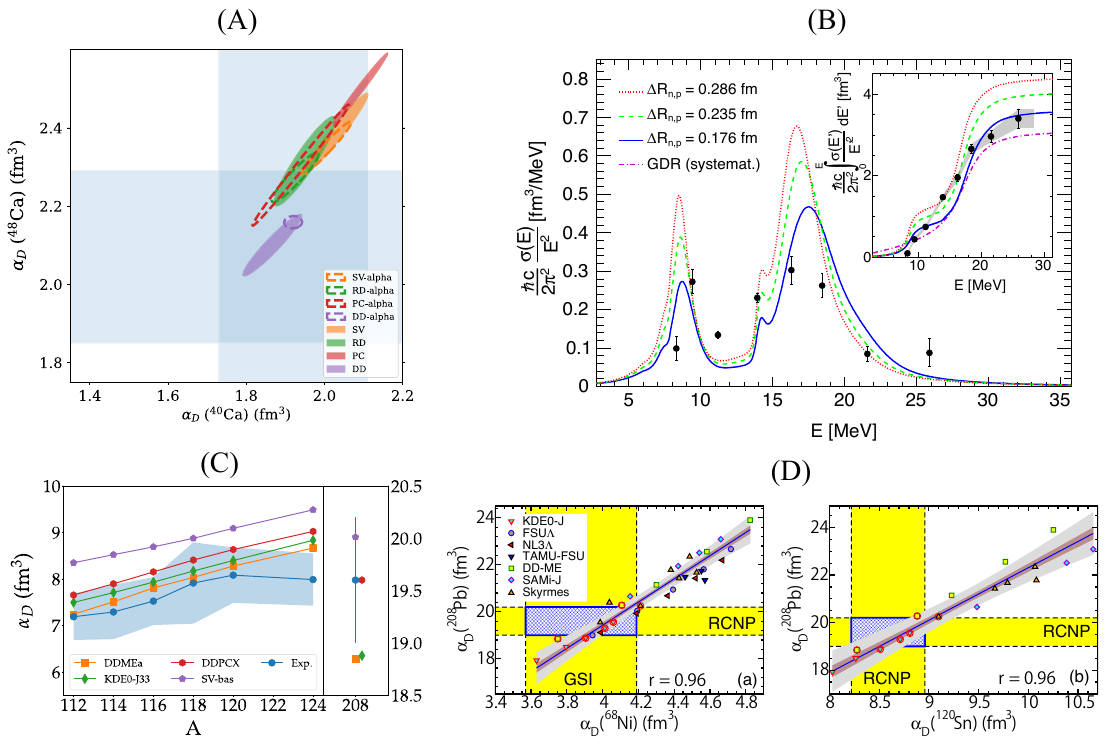}
\end{center}
\caption{
(A) Correlation of the experimental DP of $^{40}$Ca and $^{48}$Ca (blue bands) in comparison with DFT calculations without (full ellipses) and with (dashed ellipses) inclusion of the experimental DP of $^{208}$Pb \cite{tamii11} in the parameter fit.
(B) $E1$ strength distribution in $^{68}$Ni (black circles) in comparison to DFT calculations systematically varying the neutron skin thickness \cite{piekarewicz11}.
The inset show the running sum of the DP.
(C) Systematics of the DP in the stable Sn isotopes (left panel) and in $^{208}$Pb (right panel). 
The experimental values (blue dots) and their errors (blue band) are compared with DFT results from several modern interactions.
(D) Correlation (blue crossed histograms) of the DP in $^{208}$Pb with $^{68}$Ni (left panel) and $^{120}$Sn (right panel) with uncertainties (yellow bands) in comparison with DFT calculations for a large set of interactions and a linear fit with uncertainty bands.
Figures taken from (A) Ref.~\cite{fearick23}, (B) Ref.~\cite{rossi13},  Ref.~\cite{bassauer20} and (D) Ref.~\cite{rocamaza15}, where the original references can be found.
}
\label{fig:10}
\end{figure}

The DP of $^{40}$Ca and $^{48}$Ca has been studied in Refs.~\cite{fearick23} and\cite{birkhan17}, respectively.
Figure~\ref{fig:10}(A) depicts their correlation and a comparison to selected DFT results.
The four functionals are representative of widely used forms: non-relativistic Skyrme functionals SV \cite{kluepfel09} and RD \cite{erler10} with different forms of density dependence, and relativistic functionals DD \cite{niksic02} with ﬁnite-range meson-exchange coupling and PC \cite{niksic08} with point coupling. 
All four have been calibrated to the same set of ground-state data to determine the model parameters.

The predictions are displayed as filled ellipses which represent the $1\sigma$ error as defined in Ref.~\cite{reinhard21}.
The DD functional performs rather well. 
The other models tend to slightly overestimate the experimental mean values of both $^{40}$Ca and $^{48}$Ca, but their $1\sigma$ error ellipses do overlap with the experimental bands, except for PC. 
In all cases, the $\alpha_{D}$ values for both nuclei are highly correlated. 
The dashed ellipses show the effect of additionally including the experimental $\alpha_{D}$ value of $^{208}$Pb \cite{tamii11} in the fit yielding 
functionals denoted "-alpha". 
This improves the agreement with experiment and shrinks the error ellipsoids. 
The models incorporate a span of symmetry energy parameters $J = 30-35$ $(30-32)$ MeV and $L = 32-82$ $(35-52)$ MeV for the calculations excluding (including) the $^{208}$Pb data point.

The $B(E1)$ strength distribution of the unstable neutron-rich nucleus $^{68}$Ni determined in an experiment measuring Coulomb excitation in inverse kinematics \cite{rossi13} is displayed in Fig.~\ref{fig:10}(B).
The DP was extracted from a comparison to the model of Ref.~\cite{piekarewicz11}.
The model results show a sensitivity to the assumed neutron skin thickness as illustrated by the colored curves.
A value of 0.17(2) fm was extracted for the neutron skin thickness from the correlation between the two quantities.

A study of the DP in a long isotopic chain is particularly suited to investigate the connection with the neutron skin thickness.
This can be best done in the Sn isotopes with neutron numbers between 50 and 82, where the proton shell closure stabilizes the g.s.\ deformation.
There are a large number of stable isotopes and a study of the systematics of the DP was presented in Ref.~\cite{bassauer20}.
The results are summarized in Fig.~\ref{fig:10}(C) which shows the evolution of $\alpha_{\rm D}$ between mass numbers 112 and 124.
All DFT calculations predict an approximately linear increase as a function of neutron excess with roughly the same slope. 
The experimental results indicate a saturation between mass numbers 120 and 124, but the uncertainties (blue band) do not exclude a mass dependence similar to the theoretical results.
The rightmost part of Fig.~\ref{fig:10}(C) shows the predictions of the different models for the $^{208}$Pb DP after subtraction of the quasi-deuteron part (see next paragraph). 
The models closest in absolute magnitude to the data tend to underpredict $\alpha_{\rm D}$($^{208}$Pb), while those reproducing it overshoot the absolute values in the Sn chain indicating that the functionals cannot yet fully describe the mass dependence of the DP. 
We note that $E1$ strength distributions have also been measured for the unstable neutron-rich isotopes $^{130,132}$Sn \cite{adrich05} but the extracted values of $\alpha_{\rm D}$ cannot be compared directly to the results of Ref.~\cite{bassauer20}, because the experiment only provided data above neutron threshold.

Roca-Maza et al.~\cite{rocamaza15} combined the experimental DP data for $^{68}$Ni \cite{rossi13}, $^{120}$Sn \cite{hashimoto15} and $^{208}$Pb \cite{tamii11} to test a large variety of density functionals.
Since the DFT calculatons do not include contributions from the quasi-deuteron process dominating the photoabsorption cross sections above the energy region of the IVGDR, these had to be removed for a comparison \cite{rocamaza15}.
Figure~\ref{fig:10}(D) presents correlation plots between the experimental results and theoretical predictions from a wide range of DFT interactions.
Only a handful (marked in red) is capable of simultaneously describing all three data points.
Based on this reduced set, systematic predictions of $\alpha_{D}$ for other masses, $r_{skin}$ and the symmetry energy parameters could be derived.
The experimental results for $^{40,48}$Ca discussed above are fairly well described by these predictions.  

\subsection{Constraints based on ab initio models}
\label{sec:42}

\begin{figure}
\begin{center}
\includegraphics[width=\textwidth]{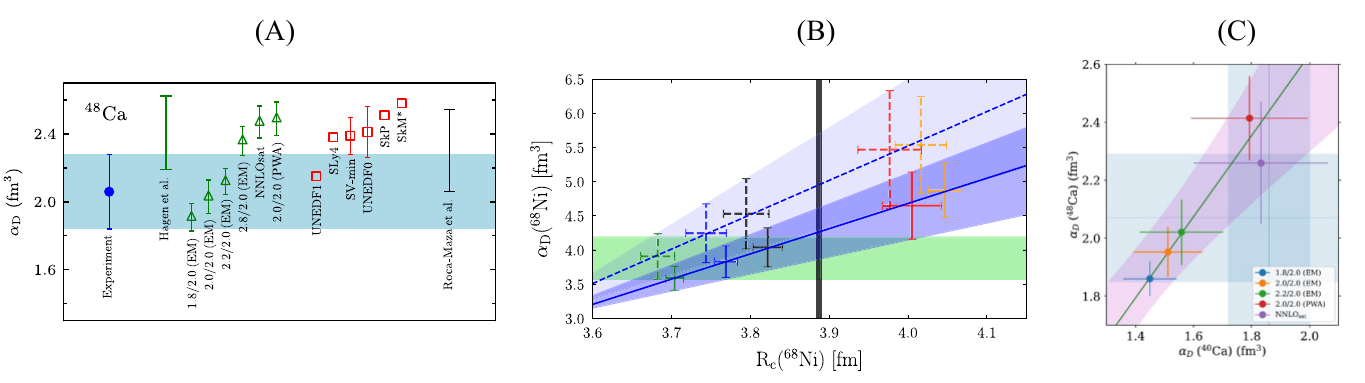}
\end{center}
\caption{
(A) Experimental DP in $^{48}$Ca (blue band) and predictions from ab initio results based on $\chi$EFT interactions (green triangles) and DFT calculations (red squares). 
The green and black bars indicate the ab initio prediction selected to reproduce the $^{48}$Ca charge radius and the range of DP predictions from Ref.~\cite{rocamaza15} simultaneously consistent
with the DP in $^{68}$Ni, $^{120}$Sn and $^{208}$Pb, cf.\ Fig.~\ref{fig:11}(D).
(B) Correlation of the experimental DP (green band) and the charge radius (black band) in $^{68}$Ni with comparison to ab initio coupled-cluster calculations up to 2p-2h (dashed crosses) and 3p-3h excitations (full crosses).
The dashed and full lines and corresponding error bands result from  linear fits to the theoretical resuts.
(C) Correlation of the experimental DP in $^{40}$Ca and $^{48}$Ca in comparison with ab initio coupled-cluster calculations including 3p-3h excitations (crosses and purple uncertainty band).
Figures taken from (A) Ref.~\cite{birkhan17} and (B)~\cite{kaufmann20}, where the original references can be found.
Figure \ref{fig:11}(C) is taken from Ref.~\cite{fearick23} but modified to include an estimate of the theoretical uncertainties shown as purple band.
\label{fig:11}
}
\end{figure}

An experimental study of the DP in $^{48}$Ca \cite{birkhan17} is of particular interest, since it is accessible for both DFT and ab initio calculations and a measurement of the neutron skin with parity-violating electron scattering is available \cite{adhikari22}.
The comparison is summarized in Fig.~\ref{fig:11}(A), where the blue band describes the experimental uncertainty. 
Ab initio results for the set of interactions from Refs.~\cite{hebeler11,ekstroem15} are displayed as green triangles and a prediction from Ref.~\cite{hagen16} based on a normalization to the $^{48}$Ca charge radius as green bar.
Results from the set of density functionals described in Ref.~\cite{hagen16} are shown as red squares with some representative error bars, and the prediction from the analysis of Ref.~\cite{rocamaza15} discussed above as black bar. 

The EDF results tend to be somewhat high compared to  experiment.
The ab initio results show a significant dependence on the chosen interaction, but it can be well approximated by a linear dependence.
This allows in principle to derive boundaries on the neutron skin thickness and the symmetry energy.
However, while the ab initio results shown were truncated in the coupled-cluster expansion at the 2p-2h level, subsequent work \cite{miorelli18} demonstrated that inclusion of 3p-3h correlations lowers the $\alpha_{\rm D}$ values by $10-20$\%.
The refined results in $^{48}$Ca are plotted in Fig.~\ref{fig:11}(C) against corresponding calculations for $^{40}$Ca \cite{fearick23}.
A high correlation similar to the DFT results shown in Fig.~\ref{fig:10}(A) is observed.
The purple uncertainty band from the ab initio results overlaps with the crossing of the experimental $1\sigma$ error bands.
In particular, the NNLOsat interaction interaction \cite{ekstroem15} accurately describing binding energies and radii of nuclei up to $^{40}$Ca as well as the saturation point of symmetric nuclear matter now reproduces both DP values. 
The importance of including 3p-3h correlations has also been demonstrated in a recent measurement of the $^{68}$Ni charge radius \cite{kaufmann20}.
Figure~\ref{fig:11}(B) illustrates the improvement in describing the correlation between the charge radius and $\alpha_{\rm D}$ \cite{rossi13} when going from the 2p-2h level (light blue band) to the inclusion of 3p-3h correlations (dark blue band).

As noted in Section~\ref{sec:22}, independent of the chosen interaction a neutron skin thickness of around 0.14 fm is predicted for $^{48}$Ca, consistent with the value deduced from the measurement of the weak form factor \cite{adhikari22}.
The simultaneous description of the data in $^{40,48}$Ca and $^{68}$Ni implies that the underlying symmetry energy parameters are correct.
A conservative estimate is provided by taking the full range of values from the set of ab initio interactions, viz.\ $J = 27 - 33$ MeV, $L = 41 -49$ MeV.

Recent work has, for the first time, been able to extend the range of ab initio DP calculations based on $\chi$EFT interactions to $^{208}$Pb \cite{hu22}.
A different technique was used to construct the interactions by history matching \cite{vernon14} using selected experimental observables in light nuclei.
Moreover, low-energy nucleon-nucleon scattering phase shifts were considered.
The latter are responsible for tight constraints to rather small values of the resulting neutron skin thickness ($0.14 - 0.20$ fm for $^{208}$Pb). 
The variation of the density dependence of the symmetry energy in these calculaitons is $L = 38 - 69$ MeV.

\subsection{Tension between polarizability and parity-violating elastic electron scattering in $^{208}$Pb}
\label{sec:43}

While in $^{48}$Ca there is fair agreement between the neutron skin thickness and symmetry energy properties derived from the different experiments, the parity-violating elastic electron scattering experiment on $^{208}$Pb \cite{adhikari21} finds a much larger neutron skin $r_{skin} = 0.28(7)$ fm than most other work.
Accordingly, an extraction of symmetry energy parameters based on the correlations established in DFT (see Sec.~\ref{sec:21}) leads to large symmetry energy values $J = 38(5)$ MeV and $L = 106(37)$ MeV in contradiction to limits derived from astrophysical observations of neutron star radii and masses as well as the tidal deformability of neutron star mergers \cite{lattimer23}.
All astrophysical constraints point towards a softer EOS.
This has led to speculations about a phase transition at intermediate densities \cite{han19}.

\begin{figure}
\begin{center}
\includegraphics[width=7cm]{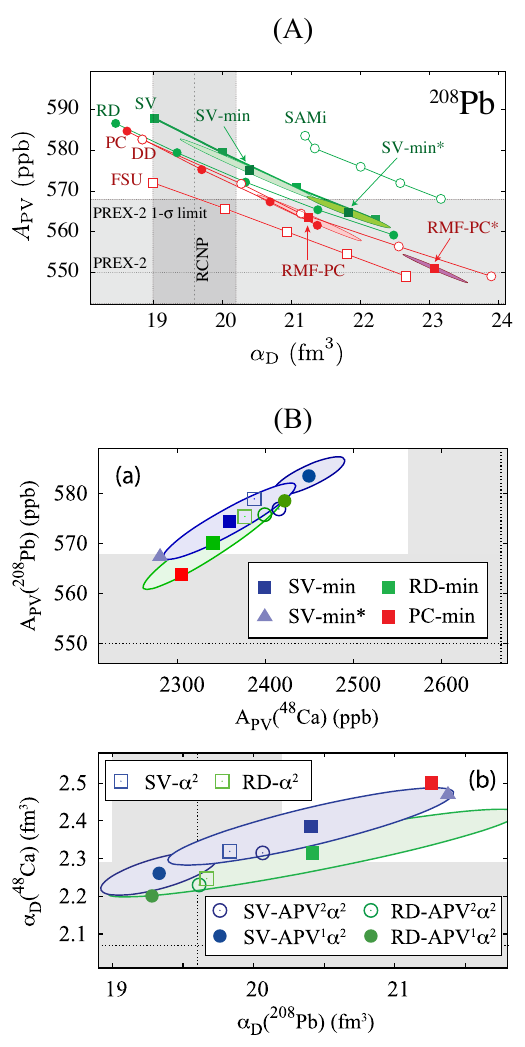}
\end{center}
\caption{
(A) Experimental parity-violating asymmetry versus DP in $^{208}$Pb (grey bands) in comparison to calculations with a set of relativistic (red) and  nonrelativistic (green) DFT interactions. Sets with systematically varied symmetry energy $J$ are connected by lines. 
Representative $1\sigma$ error ellipses are shown for the interaction indicated by squares.
Figure taken from Ref.~\cite{reinhard21}, where the original references can be found.
(B) Correlation of experimental parity-violating asymmetries (top) and DP (bottom) in $^{48}$Ca and $^{208}$Pb (grey bands) in comparison a set of DFT interactions.
Representative $1\sigma$ error ellipses are shown for the interaction indicated by squares.
Figure taken from Ref.~\cite{reinhard22}, where the original references can be found.
}
\label{fig:12}
\end{figure}

Because of the strong correlation between $r_{\rm skin}$  and $\alpha_{\rm D}$ for a given nucleus and $r_{\rm skin}$ values of different nuclei in DFT models, Reinhard et al.~\cite{reinhard21,reinhard22} investigated whether it is possible to construct a DFT interaction capable of simultaneously describing the data for $^{48}$Ca and $^{208}$Pb.
The analysis was based on representative families of non-relativistic and relativistic functionals.
The isovector properties of EDFs are typically not well constrained by the input data used to fit the model parameters. 
As illustrated in Fig.~\ref{fig:12}(A) for the case of $^{208}$Pb, it is possible to vary the symmetry energy parameters -- and thereby the predicted $r_{skin}$ and $\alpha_{D}$ -- over a fairly large range maintaining comparable description of ground-state properties \cite{reinhard21}.
Figure~\ref{fig:12}(B) \cite{reinhard22} demonstrates that the polarizabilities and the neutron skin thickness of $^{48}$Ca could be consistently described, but it was impossible to construct an EDF simultaneously accounting for the neutron skin thickness of $^{208}$Pb extracted from the PREX experiment \cite{adhikari22}.
Similar conclusions were drawn in Refs.~\cite{piekarewicz21,yueksel22}.
In another recent attempt \cite{reed24}, a DFT interaction reasonably accounting for the measured parity-violating asymmetries in both the PREX and CREX experiments was constructed, but at the expense of unusual properties of the symmetry energy curvature and a very strong isovector coupling leading to density fluctuations in the nuclear interior. 

\subsection{Volume and surface contributions to the symmetry energy}
\label{sec:44}

\begin{figure}
\begin{center}
\includegraphics[width=8cm]{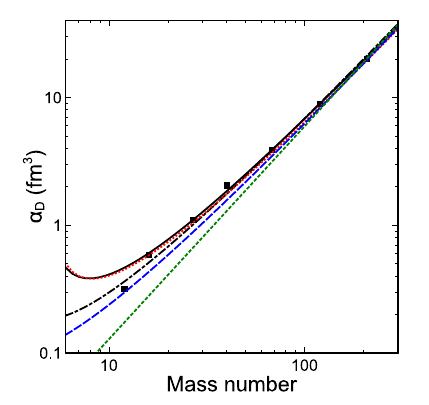}
\end{center}
\caption{
Experimental DP for a set of nuclei as a function of mass number (full squares).
The green and blue lines are fits with original Migdal model (Eqs.~(1) and (2) in Ref.~\cite{orce15}).
The black lines are fits allowing for a surface term of the symmetry energy including (dashed-dotted) and excluding (full) the data point for $^{12}$C. 
The red line shows a fit with the prediction of Ref.~\cite{steiner05} using the "$\mu_n$" approach. 
Figure taken from Ref.~\cite{vonneumanncosel16}, where the original references can be found.
}
\label{fig:13}
\end{figure}

Another way of extracting properties of the symmetry energy is a study of the mass dependence of the DP.
A simple power law $\sigma_{-2} \propto A^{5/3}$ based on a model of two interpenetrating fluids has been given by Migdal, where $\sigma_{-2}$ denotes the second inverse moment of the photoabsorption cross sections and $\sigma_2 \simeq \alpha_{\rm D}$ in units of mb/MeV (Ref.~\cite{orce20} and Refs.\ therein).
A proportionality constant $2.4\times 10^{-3}$ has been determined by Orce \cite{orce15} from a fit to $(\gamma,xn)$ data over a wide mass range.
Figure \ref{fig:13} \cite{vonneumanncosel16} shows a comparison with a combined data set of $\alpha_{D}$ measurements in light nuclei \cite{ahrens75} with then available (2016) data from relativistic Coulomb excitation for heavier nuclei as green short-dashed line. 
Note that results for $A < 12$ from Ref.~\cite{ahrens75} are neglected because the hydrodynamical picture
is highly questionable and corrections due to the magnetic polarizability are large \cite{knuepfer81} for
these very light nuclei.
Results are severely underestimated in lighter nuclei where charged-particle decay dominates.
The mass dependence is reasonably described for larger masses but the proportionality coefficient of Ref.~\cite{orce15} is too low, since additional contributions from the strength below neutron threshold, as discussed in Sec.~\ref{sec:33}, need to be considered.

Furthermore, for masses $A \leq 40$ surface contributions have to be considered modifying the volume term of the symmetry energy dominating for heavy nuclei. 
These can be parameterized as \cite{vonneumanncosel16}
\begin{equation}
\sigma_{-2} = \frac{0.0518 \, A^2}{S_v (A^{1/3} - \kappa)} \, {\rm mb/MeV}.
\label{eq:surface}
\end{equation}
Here $\kappa = S_s/S_v$, and $S_s$ and $S_v$ denote the surface and volume coefficients of the symmetry energy, respectively.
The numerical coefficient is obtained from Migdal’s approach.
A fit with $S_s,S_v$ parameters from binding energies of isobaric nuclei \cite{tian14} shown in Fig.~\ref{fig:13} as blue long-dashed line still underestimates the lower-mass data.
Parameters of the study of Ref.~\cite{steiner05} provide a better description (dotted red line).
Results of a free fit of Eq.~(\ref{eq:surface}) crucially depend on the inclusion (dotted-dashed black line) or exclusion (solid black line) of the $^{12}$C data point.
The latter provides a better fit with $S_v = 25.6(8)$ MeV, $\kappa = 1.66(5)$ \cite{vonneumanncosel16} close to Ref.~\cite{steiner05}.
$S_v$ can be interpreted as $J$, but measured at about $2/3$ of the saturation density \cite{centelles09,rocamaza13}.

\section{Conclusions and outlook}
\label{sec:5}

We present a review of methods to measure the isovector $E1$ response in nuclei and the extraction of the dipole polarizability from these data.
The discussion focuses on recent results obtained with inelastic proton scattering under extreme forward angles at RCNP.
At energies of a few hundred MeV, relativistic Coulomb excitation dominates the cross sections in these kinematics. 
The method combines certain advantages compared to other experimental techniques: (i) it measures the absorption and is thus independent of the knowledge of branching ratios, (ii) a separation of $E1$ and $M1$ contributions to the cross sections can be achieved with different independent approaches, and (iii) the relevant excitation energy region from well below the neutron threshold across the IVGDR can be covered in a single experiment.

Constraints on the neutron skin thickness of nuclei and the parameters of the symmetry energy can be extracted from the strong correlations between these three quantities seen in all microscopic models.
Results from nuclei covering a mass range between $^{40}$Ca and $^{208}$Pb consistently favor small neutron skins and a soft density dependence of the EOS around saturation density. 
In $^{208}$Pb serving as a benchmark for theory, this finding is at variance with the results of a parity-violating elastic electron scattering experiment (PREX), while a similar study of $^{48}$Ca (CREX) conforms.
The PREX result, hard to interpret in the framework of present theory, has led to an initiative (called MREX) for a study with improved statistical and systematic errors at the new Mainz high-current MESA accelerator \cite{schlimme04}.

While the mass dependence of the DP is reasonable covered by the available data, future work should explore other degrees of freedom like the variation of neutron excess along isotopic chains and the role of deformation.
Furthermore, the experimental uncertainties of the DP for key nuclei can be improved by the availability of independent measurements as illustrated in Fig.~\ref{fig:6}.
New high-brilliance LCBS photon beam facilities are under construction at the Extreme Light Infrastructure - Nuclear Physics (ELI-NP) in Bucarest \cite{gales18,tanaka20} and the Shanghai Laser Electron Gamma Source (SLEGS) at the Shanghai Synchrotron Radiation Facility \cite{wang22}.
Combined with advanced techniques for neutron detection \cite{gheorghe21}, these facilities promise a new quality of precision for $(\gamma,xn$) experiments. 

Major steps can be expected in the future at radioactive ion beam facilities providing access to cases with much larger neutron excess than achievable for stable nuclei.
Experimental tools for the measurement of relativistic Coulomb excitation in reverse kinematics are available, and pioneering studies of the dipole response in unstable nuclei have been performed at GSI \cite{adrich05,wieland09,rossi13}.
First results for the neutron-rich isotope $^{52}$Ca investigated at RIKEN have been reported \cite{togano24}.
Because of the high energy/nucleon available, future experiments at FAIR are particularly promising for research on the dipole polarizability of exotic neutron-rich nuclei \cite{aumann24}.

\section*{Conflict of Interest Statement}

The authors declare that the research was conducted in the absence of any commercial or financial relationships that could be construed as a potential conflict of interest.

\section*{Author Contributions}

The layout of the article was designed by both authors. PvNC prepared a draft of most of the text while AT prepared the text in Sec.~3.1 and the figures. 
The final text was reviewed by both authors.

\section*{Funding}
The author(s) declare that financial support was received for
the research, authorship, and/or publication of this article.
This work was supported by the Deutsche Forschungsgemeinschaft (DFG, German Research Foundation) under Contract No.\ SFB 1245 (Project ID No.\ 79384907), by the Research Council of Norway through its grant to the Norwegian Nuclear Research Centre (Project No. 341985), by the JSPS KAKENHI Grant Number 25H00641, and the Japan-South Africa Bilateral Funding Grant Number JPJSBP 120246502.

\section*{Acknowledgments}

PvNC thanks the nuclear physics group at the University of Oslo for their kind hospitality during a stay where major parts of this work were done. 


\bibliographystyle{Frontiers-Harvard} 

\bibliography{References}

\end{document}